\documentclass[10pt,conference]{IEEEtran}
\IEEEoverridecommandlockouts
\usepackage{cite}
\usepackage{amsmath,amssymb,amsfonts}
\usepackage{algorithmic,comment}
\usepackage{graphicx,cleveref}
\usepackage{subcaption}
\usepackage{textcomp}
\usepackage{xcolor}
\usepackage{url}
\def\BibTeX{{\rm B\kern-.05em{\sc i\kern-.025em b}\kern-.08em
    T\kern-.1667em\lower.7ex\hbox{E}\kern-.125emX}}

\begin{document}

\title{Collaborative P4-SDN DDoS Detection and Mitigation with Early-Exit Neural Networks \\

\thanks{This research has received funding from the European Union’s Horizon research and innovation program under grant agreement No 101070374.}
}

\author{\IEEEauthorblockN{Ouassim Karrakchou$^{{\star}}$, Alaa Zniber$^{{\star}}$, Anass Sebbar$^{{\star}}$, Mounir Ghogho$^{{\dagger}{\ddagger}}$}

\IEEEauthorblockA{$^{\star}$ \textit{TICLab, International University of Rabat, Morocco}}
\IEEEauthorblockA{$^{\dagger}$ \textit{College of Computing, University Mohammed VI Polytechnic, Morocco}}
\IEEEauthorblockA{$^{\ddagger}$ \textit{School of Electronic and Electrical Engineering, University of Leeds, UK}} 
\{ouassim.karrakchou, alaa.zniber, anass.sebbar\}@uir.ac.ma, mounir.ghogho@um6p.ma}

\makeatletter
\def\ps@IEEEtitlepagestyle{%
  \def\@oddfoot{\mycopyrightnotice}%
  \def\@oddhead{\hbox{}\@IEEEheaderstyle\leftmark\hfil\thepage}\relax
  \def\@evenhead{\@IEEEheaderstyle\thepage\hfil\leftmark\hbox{}}\relax
  \def\@evenfoot{}%
}
\def\mycopyrightnotice{%
  \begin{minipage}{\textwidth}
  \centering \scriptsize
  ~\copyright~2025 IEEE.  Personal use of this material is permitted.  Permission from IEEE must be obtained for all other uses, in any current or future media, including reprinting/republishing this material for advertising or promotional purposes, creating new collective works, for resale or redistribution to servers or lists, or reuse of any copyrighted component of this work in other works.
  \end{minipage}
}
\makeatother

\maketitle

\begin{abstract}
Distributed Denial of Service (DDoS) attacks pose a persistent threat to network security, requiring timely and scalable mitigation strategies. In this paper, we propose a novel collaborative architecture that integrates a P4-programmable data plane with an SDN control plane to enable real-time DDoS detection and response. At the core of our approach is a split early-exit neural network that performs partial inference in the data plane using a quantized Convolutional Neural Network (CNN), while deferring uncertain cases to a Gated Recurrent Unit (GRU) module in the control plane. This design enables high-speed classification at line rate with the ability to escalate more complex flows for deeper analysis. Experimental evaluation using real-world DDoS datasets demonstrates that our approach achieves high detection accuracy with significantly reduced inference latency and control plane overhead. These results highlight the potential of tightly coupled ML-P4-SDN systems for efficient, adaptive, and low-latency DDoS defense.
\end{abstract}

\begin{IEEEkeywords}
P4, SDN, DDoS, Early-Exit Neural Networks
\end{IEEEkeywords}

\section{Introduction}
Distributed Denial of Service (DDoS) attacks remain a critical threat to the availability and reliability of online services. By overwhelming targeted resources with high volumes of malicious traffic from distributed sources, these attacks can severely degrade network performance, disrupt service continuity, and inflict substantial financial and reputational damage. The widespread adoption of cloud-native architectures, proliferation of Internet of Things (IoT) devices, and increased reliance on high-speed networks have collectively expanded the attack surface and exacerbated the severity and frequency of such attacks \cite{chaudharyDDoSAttacksIndustrial2023}.

Efficiently detecting and mitigating DDoS attacks—particularly in real-time—continues to be a pressing research and operational challenge. Conventional mitigation techniques, which often rely on static rules or threshold-based detection, are inadequate in addressing dynamic and sophisticated attack vectors \cite{pengSurveyNetworkbasedDefense2007}. Furthermore, many existing solutions depend heavily on centralized processing or offline analysis, resulting in response delays and scalability limitations, especially during high-volume attacks \cite{liComprehensiveSurveyDDoS2023}.

In recent years, machine learning (ML) techniques have gained significant traction for DDoS detection, offering the ability to learn traffic patterns and identify anomalies that traditional rule-based systems fail to capture \cite{survey_dl_ddos}. However, most of these methods are used to process offline packet traces, thus limiting their applicability to real-time contexts. Simultaneously, Software-Defined Networking (SDN) has emerged as a powerful paradigm for network programmability, providing centralized control, global visibility, and dynamic policy enforcement, which are well-suited to the needs of DDoS mitigation \cite{kreutzSoftwaredefinedNetworkingComprehensive2015}. The integration of ML-based detection mechanisms within the SDN control plane has yielded promising results in identifying and responding to attack traffic \cite{swamiSoftwaredefinedNetworkingbasedDDoS2019}. However, these approaches typically perform inference outside the data plane, introducing latency and creating potential bottlenecks under high-traffic conditions.

The emergence of programmable data planes, notably through the P4 language \cite{bosshartP4ProgrammingProtocolindependent2014}, opens new avenues for deploying custom packet processing logic directly at the network edge. Recent studies have explored the feasibility of implementing ML models within the data plane to enable low-latency detection \cite{zhangMachineLearningBasedToolbox2024}. Nevertheless, current approaches fall short of fully leveraging the potential of collaborative designs that integrate both P4-based data planes and SDN controllers. Existing literature either limits the role of the data plane \cite{wangCollaborativeDefenseHybrid2024} or SDN controller \cite{liComprehensiveSurveyDDoS2023} to non-ML-based reactive methods,  uses the controller solely for offline training, without dynamic collaboration during runtime \cite{zhangTwoStageConfidenceBasedIntrusion2023}, or deploys the ML models completely either in the data or control planes despite allowing online collaboration between those two planes \cite{doriguzzi-corinIntroducingPacketlevelAnalysis2024, musumeciMachineLearningEnabledDDoSAttacks2021}.

In this paper, we present a novel collaborative architecture for DDoS detection and mitigation that integrates P4-programmable data plane elements with an SDN control plane in a synergistic manner. At the core of our approach is the split deployment of an early-exit dynamic neural network, which allows for the inference to be terminated at an intermediate layer if the model reaches a high-confidence prediction, reducing computation and latency. The proposed network architecture comprises a lightweight Convolutional Neural Network (CNN) frontend that processes individual packets to extract salient features, and a Gated Recurrent Unit (GRU) backend that captures temporal dependencies across packet sequences for more nuanced detection. To enable early exiting, we introduce an intermediate classifier positioned between the CNN frontend and the GRU backend. This module assesses whether the features extracted by the CNN are sufficient to make a high-confidence prediction. If so, inference is terminated early, and mitigation is triggered.

During deployment, the network is split across the data and control planes: the CNN frontend and intermediate classifier are quantized and implemented in P4 using integer arithmetic, allowing them to operate at line rate, while the GRU backend is hosted in the SDN controller. This architectural design offers a flexible trade-off between inference complexity and response time. When the CNN is able to confidently classify a flow as malicious, mitigation is executed autonomously in the data plane, thereby minimizing latency and reducing load on the control plane. Conversely, in cases where the CNN's confidence is insufficient, a compact feature vector derived from the CNN is forwarded to the controller for continued inference using the GRU, thus allowing for a more in-depth analysis in the SDN controller. This collaborative inference strategy harnesses the responsiveness of in-network processing while retaining the analytical depth of centralized computation, ultimately enabling fast, scalable, and adaptive DDoS mitigation.

The main contributions of this paper are as follows:
\begin{enumerate}
    \item We propose a collaborative P4-SDN architecture for DDoS detection and mitigation, which distributes inference between the data and control planes to combine fast in-network processing with centralized intelligence.
    \item We design and implement a split early-exit neural network, consisting of a quantized CNN deployed in the data plane and a GRU module in the SDN controller. This architecture enables partial inference and adaptive decision-making based on model confidence.
    \item We evaluate our system using real-world DDoS datasets, demonstrating that our approach achieves high detection accuracy with low inference latency and reduced control plane overhead.
\end{enumerate}

The rest of this paper is organized as follows. Section II reviews related work on DDoS detection and early exit neural networks. Section III details the proposed architecture and its components. Section IV presents the experimental evaluation, and Section V concludes the paper.

\section{Related Work}

This section surveys two key areas relevant to our contributions: early-exiting neural networks and learning-based methods for DDoS detection. Our work bridges these domains through a novel split architecture for collaborative inference within programmable networks.

\subsection{Early-Exiting Neural Networks}

Early-exiting neural networks are designed to reduce computational overhead by allowing intermediate layers to make confident predictions and terminate inference early~\cite{early_exit_survey}. This paradigm was initially introduced for image classification~\cite{shallow_deep}, where shallow classifiers are embedded at various depths of the network to make early decisions when the output is sufficiently reliable. Subsequent extensions have adapted this concept to other domains, including object detection~\cite{obj_det_early_exit}, natural language processing~\cite{early_exit_nlp}, and speech enhancement~\cite{nsnet2}.

The early-exit mechanism is closely related to split computing, which seeks to partition deep models across heterogeneous platforms—typically deploying shallow components on edge devices and deeper components in the cloud~\cite{split_early_survey}. Collaborative training of such split models is critical to prevent performance degradation due to independently optimized components. Two main training strategies dominate the literature: Layer-wise Training (LT) and Joint Training (JT). LT trains the exits sequentially by freezing earlier submodels, but its greedy nature can lead to suboptimal global performance. This has led to techniques such as model ensembling to mitigate its limitations~\cite{ boosted_eenn}. JT, on the other hand, uses multi-objective optimization to jointly minimize the losses from all exits—often via linear scalarization~\cite{deep_feature_surgery}.

While these methods address the computational cost of deep inference, most existing work is limited to traditional computing environments (e.g., edge-cloud systems) and does not address the challenges of deployment in data-plane network devices. In contrast, we propose a collaborative P4-SDN architecture that incorporates early exiting directly within the programmable data plane.

\subsection{Learning-Based Methods for DDoS Detection}

DDoS detection has been extensively studied as a classification problem using machine learning techniques. Classical models such as random forests and XGBoost have been applied successfully to known attack signatures~\cite{ddos_ml_models}, but they often lack the adaptability required to respond to evolving attack patterns in real-time.

Recent advances leverage deep learning models for their superior representational power. These approaches are typically categorized based on their handling of temporal structure in network traffic. Instance-based models, including multi-layer perceptrons and CNNs, have shown promise in learning spatial patterns in pre-processed flow features (e.g., TCP flag counts) ~\cite{ddos_cnn}. More sophisticated architectures, such as autoencoders~\cite{autoencoder} and transformers~\cite{transformer}, have been used for anomaly detection and feature abstraction.

Another body of work exploits temporal dependencies between packets or flow feature states through Recurrent Neural Networks (RNNs) and their variants, such as GRUs and LSTMs~\cite{gru_based}, which have demonstrated robustness in dynamic traffic environments. Hybrid architectures have also emerged, combining CNNs for spatial filtering and RNNs for sequential modeling. For example,\cite{sequential_lstm_cnn} proposes a layered model integrating both components, while\cite{two_lines_cnn_lstm} introduces a dual-path strategy to process inputs through parallel convolutional and recurrent modules.

While these studies focus primarily on detection accuracy and model improvements, they typically assume a centralized inference pipeline and overlook deployment constraints in real-time high-speed networks. Our work addresses this gap by proposing a distributed DDoS detection framework that partitions inference between the data and control planes. Unlike existing approaches, our design enables adaptive, confidence-based early exits at the P4 switch to minimize control plane load, while preserving high detection accuracy through deeper inference at the controller when necessary.

\section{Proposed Architecture}

\begin{figure}[t]
    \centering
    \includegraphics[width=0.95\linewidth]{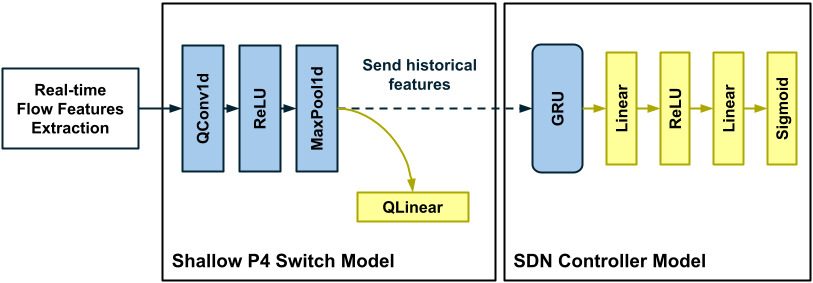}
    \caption{Overview of the proposed architecture}
    \label{fig:model}
    \vspace{-5mm} 
\end{figure}

This section presents the design of our collaborative P4-SDN framework for DDoS detection, detailing its key components, their interactions, and the flow of data across the data and control planes.

The primary goal of our architecture is to enable low-latency, high-accuracy detection of DDoS attacks by distributing inference across the programmable data plane and the SDN control plane. To achieve this, we introduce a split neural network with early-exit capabilities, allowing the system to handle straightforward cases directly in the switch, while deferring ambiguous or low-confidence decisions to the controller for deeper analysis. This collaborative mechanism reduces unnecessary control plane interactions and allows rapid, in-network mitigation of attacks.

An overview of the proposed architecture is shown in Fig.~\ref{fig:model}. The system consists of the following main components:

\begin{itemize}
    \item \textbf{Real-time flow feature extraction:} Conducted within the P4 switch, this component identifies network flows and computes lightweight flow statistics and aggregates them into a fixed-size feature vector representing flow behavior.
    
    \item \textbf{Shallow quantized CNN with early exit:} The extracted features are processed by a shallow, quantized CNN deployed in the data plane. An early-exit classifier attempts to classify each flow with minimal computation. If confidence is high, a decision is made locally at the switch.
    
    \item \textbf{Data-to-control plane communication:} In cases of low confidence, the CNN’s last 10 feature maps are encapsulated in a custom packet and sent to the SDN controller. This approach minimizes load on the control plane southbound interface.
    
    \item \textbf{GRU-based controller inference:} At the SDN controller, a GRU network processes the received feature maps, performing deeper temporal analysis and leveraging global context to reach a final classification. The output of the GRU results in an action for the flow (drop, allow, or send to the controller) that is transmitted to the P4 switch for enforcement.
\end{itemize}

This distributed inference framework combines the speed of in-network processing with the depth of centralized intelligence, enabling scalable and adaptive DDoS detection in programmable network environments. The next parts provide more details on the different components of our proposed architecture.

\subsection{Real-time flow feature extraction in P4}

In our architecture, flow-based detection begins with the identification and tracking of individual flows within the P4 switch. Flows are defined using the standard 5-tuple: source IP address, destination IP address, transport protocol, source port, and destination port. To ensure correct flow directionality handling, forward and backward TCP flows (e.g., from client to server and server to client) are treated as part of the same bidirectional flow entry.

We compute a subset of flow features in real-time inspired by the CICFlowMeter tool\footnote{\url{https://github.com/CanadianInstituteForCybersecurity/CICFlowMeter}}. Due to the computational and memory constraints of P4 data plane processing, we selectively include only those features that can be efficiently calculated using packet header fields or simple arithmetic operations over stateful flow data. This includes metrics such as packet header length, inter-arrival times (IAT), and TCP flag counts. Conversely, we exclude features requiring complex operations or significant historical state, such as means or standard deviations over the entire flow. This design trade-off results in a set of 28 flow features that can be accurately computed in real-time. To maintain per-flow state across packets, the unnormalized values of the last computed features for each flow are stored in a P4 hash table implemented using stateful P4 registers.

Prior to inference by the shallow CNN, each feature is normalized using Min-Max scaling based on the range observed during offline training. Specifically, each feature \( f_i \) is transformed into an unsigned 8-bit integer using the following quantization formula:

\begin{equation}
    f_i^{\text{norm}} = 127 \cdot \left\lfloor \frac{f_i - \min f_i}{\max f_i - \min f_i} \right\rfloor
\end{equation}

This quantization enables seamless integration with the quantized CNN model deployed in the data plane.

\subsection{Mixed-Precision Early-Exit CNN-GRU Model}

To support collaborative and adaptive inference across the data and control planes, we design a mixed-precision neural architecture consisting of a quantized CNN deployed within the P4 switch and a full-precision GRU module running on the SDN controller. This model, illustrated in \Cref{fig:model}, is explicitly tailored to enable early exits and minimize inference latency and overhead in programmable network environments.

The CNN module, executed in the data plane, comprises a 1D convolutional layer with 16 filters, a kernel size of 3, a stride of 1, and zero padding. This is followed by a ReLU activation and a max-pooling layer outputting a single value per channel. A final linear classification layer is appended to facilitate early decision-making directly in the switch. All computations in this component are quantized to 8-bit integers, adhering to the hardware limitations of the P4 pipeline. To ensure compatibility with integer-only arithmetic, the quantization procedure follows the techniques introduced by Jacob et al.~\cite{jacobQuantizationTrainingNeural2018}, enabling efficient and accurate fixed-point inference.

Let $\hat{y}_\text{P4}$ denote the output of the quantized linear classifier. In a standard early-exit setting, this output would typically be passed through a sigmoid function and compared against two confidence thresholds, $\tau_{\text{benign}}$ and $\tau_{\text{attack}}$. However, due to the lack of floating-point support in P4, we instead use the inverse of the sigmoid function to implement the thresholding logic. The classification decision at the switch is therefore based on precomputed integer thresholds derived from the logit transformation (sigmoid inverse):

\begin{equation*}
    \text{P4 Switch Decision} = \begin{cases}
        \text{Benign}, & \text{if} \ \hat{y}_\text{P4} <  \ln\left(\frac{\tau_{\text{benign}}}{1 - \tau_{\text{benign}}}\right) \\
        \text{Attack}, & \text{if} \ \hat{y}_\text{P4} > \ln\left(\frac{\tau_{\text{attack}}}{1 - \tau_{\text{attack}}}\right)
    \end{cases}
\end{equation*}

Since $\ln\left(\frac{\tau}{1 - \tau}\right)$ is constant for a given threshold $\tau$, it can be precomputed offline and encoded as an 8-bit integer, ensuring compatibility with the data plane's computational constraints.

When the output $\hat{y}_\text{P4}$ lies within the range defined by the two confidence thresholds—indicating insufficient certainty for a definitive classification—the decision is deferred to the SDN controller. In such cases, the current output of the max-pooling layer is concatenated with those of the preceding nine packets, forming a sequence of ten feature vectors. These vectors are retrieved from a circular buffer maintained in the P4 switch using registers and are encapsulated in a custom packet for transmission to the controller. Upon reception, the controller dequantizes the sequence and processes it using a GRU network. To ensure consistent inference behavior across flows, the GRU's hidden state is reinitialized to zero for each incoming sequence. Operating in a 64-dimensional latent space, the GRU captures temporal dependencies among the packet-level features. The final hidden state is then passed through a two-layer fully connected classifier with ReLU activation, where the intermediate layer is sized to half the input dimension. This controller-side model operates in full floating point precision and serves to produce a refined classification decision when the P4 model's confidence is insufficient.

Prior to deployment, the entire architecture is trained end-to-end using Quantization Aware Training (QAT). The optimization objective is defined as the weighted sum of the binary cross-entropy losses from both the switch-level and controller-level classifiers. Equal weighting is used to ensure that both branches of the model contribute effectively to the learning process, enabling confident early exits without degrading overall detection accuracy.

\subsection{P4-SDN Communication}

The communication protocol between the data and control planes is designed to prioritize localized decision-making at the P4 switch, thereby maximizing in-network detection and mitigation of DDoS attacks. Control plane involvement is reserved for uncertain cases only, in order to reduce processing delays and minimize the load on the southbound interface. To this end, when the P4 switch encounters low-confidence outputs from the early-exit CNN classifier, it transmits only the max-pooling feature maps—representing a compact, high-level summary of the input flow's behavior—rather than raw packet data or full feature vectors. This selective communication strategy ensures efficient bandwidth utilization and reduced controller overhead.

At the controller, the GRU-based classifier processes the received feature sequence and produces an output that is evaluated against two predefined confidence thresholds, analogous to those used in the P4 classifier. Based on the GRU output, one of three actions is determined: (i) \textit{allow}, for benign flows; (ii) \textit{drop}, for attack flows; or (iii) \textit{notify}, for uncertain flows. The resulting action is sent back to the P4 switch, where it is installed as a forwarding rule for future packets belonging to the same flow. This mechanism ensures that once a classification decision is made by the controller, subsequent packets are handled entirely in the data plane without repeated controller intervention.

To further enhance system robustness, we leverage the controller's more expressive GRU-based model to validate the CNN classifier’s decision over time. Specifically, if a given flow reaches a threshold of 500 packets, a re-evaluation is triggered in which the controller re-checks the switch-level decision using its GRU. Furthermore, even with a control-plane rule installed, flows are reverified periodically by the GRU. These verification mechanisms mitigate the risk of misclassification by the less expressive, quantized CNN, and ensure that evolving or long-lived flows are periodically reassessed using the controller's full-precision model.

\section{Experiments}
To evaluate the effectiveness of our proposed P4-SDN collaborative framework for DDoS detection, we conduct two complementary experiments. The first is a performance evaluation of our mixed-precision early-exit CNN-GRU model using the CIC-DDoS 2019 dataset \cite{cicddos}, a widely adopted benchmark for DDoS detection that captures realistic traffic patterns. The dataset contains labeled network traffic collected over two days (November 3rd and December 1st), encompassing a variety of benign flows interleaved with 12 distinct types of TCP- and UDP-based DDoS attacks. This experiment focuses on the detection accuracy and the early-exit ratio across varying confidence thresholds, thereby quantifying the trade-offs between local (switch-level) and deferred (controller-level) inference.

The second experiment is a network simulation using NS-3, in which we integrate our full architecture 
into a realistic network scenario. DDoS attack traffic is replayed from the CIC-DDoS 2019 dataset to assess the real-time responsiveness, scalability, and communication overhead of our framework under operational conditions. This simulation demonstrates the system's ability to rapidly detect and mitigate DDoS flows in a programmable network environment while minimizing control plane intervention.

\subsection{Early-Exit CNN-GRU Model Performance} 

We evaluate our mixed-precision early-exit CNN-GRU model using the CIC-DDoS-2019 dataset. For training and validation, we use data from December 1st. The model is trained using the Adam optimizer with an initial learning rate of 0.001 and a weight decay of 0.0005. A learning rate scheduler reduces the rate upon validation loss stagnation (patience = 5 epochs), and early stopping is applied if the validation loss fails to improve over 10 consecutive epochs. To address class imbalance, we adopt a weighted binary cross-entropy loss where the positive (attack) class is weighted proportionally to the ratio of benign to attack samples.

\begin{table}[htbp]
    \centering
    \caption{Early Exit Performance on Test Data}
    \begin{tabular}{c|c|c}
    \hline
        \textbf{Exit Point} & \textbf{Precision} & \textbf{F1-score} \\
        \hline
         P4 Switch & INT8 & 99.91 \\
         \hline
         SDN Controller & FP32 & 99.90 \\
         \hline
    \end{tabular}
    
    \label{tab:perf_ee}
\end{table}

Table~\ref{tab:perf_ee} presents the F1-scores achieved at both the early (P4 switch) and final (SDN controller) exit points. Despite the model's structural constraints—namely the joint optimization of two exit branches under INT8 quantization constraints—we achieve similar performance compared to state-of-the-art deep sequential architectures previously proposed for the CIC-DDoS-2019 dataset~\cite{survey_dl_ddos}.

\begin{figure}[t]
    \centering
    \includegraphics[width=0.85\linewidth]{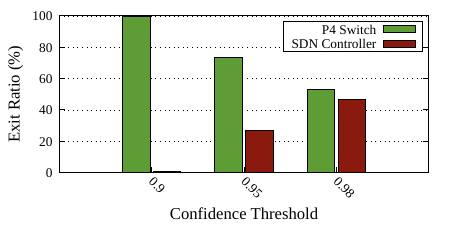}
    \caption{Exit Ratios for Different Thresholds on Test Data}
    \label{fig:ratios}
    \vspace{-5mm} 
\end{figure}

To quantify the effectiveness of early exits in reducing controller load, we analyze the distribution of exit points across different confidence thresholds $\tau = \tau_{\text{attack}} = 1-\tau_{\text{benign}}$. Figure~\ref{fig:ratios} shows the proportion of samples exiting at the P4 switch and at the controller. Notably, a majority of test flows are confidently classified at the switch level under a 90\% confidence threshold, significantly reducing the volume of data transmitted to the control plane. As expected, increasing the threshold leads to a reduction in switch-side exits due to the shallow nature and reduced precision of the in-network CNN. This illustrates the trade-off between classification confidence and communication efficiency in early-exit designs.

\subsection{Network-Level Evaluation in NS-3 Simulation}

To evaluate the practical performance of our complete architecture in a realistic network environment, we implemented the full collaborative inference pipeline on our previously proposed EP4 programmable switches \cite{karrakchouEP4ApplicationawareNetwork2021}, integrated within the NS-3 network simulator. The simulation topology consists of two EP4 switches, S1 and S2, connected in a linear topology. Each switch is connected to two end-host nodes: S1 to C1 and C2, and S2 to C3 and C4. Additionally, both switches maintain control-plane connections to a centralized SDN controller node. All network links are configured with a bandwidth of 1.5~Mbps and a propagation delay of 10~ms.

The simulation runs for a total duration of 30~seconds. At \(t = 0\)~s, a benign TCP flow is initiated from C1 to C4, with C1 uploading data at a maximum rate of 2~Mbps. At \(t = 10\)~s, the first DDoS attack begins from C2 targeting C4. This attack comprises multiple DDoS flows starting at various times extracted from PCAP traces of the November 3rd capture day in the CIC-DDoS 2019 dataset~\cite{cicddos}, ensuring realistic traffic characteristics. At \(t = 20\)~s, a second DDoS attack is launched from C3 to C4. This attack is a SYN flood generated using the Scapy packet crafting tool, allowing us to test the generalization ability of our detection model on traffic patterns not present in the CIC-DDoS dataset.

\begin{figure*}[t]
    \centering
    \begin{subfigure}[b]{0.44\textwidth}
        \centering
        \includegraphics[width=\linewidth]{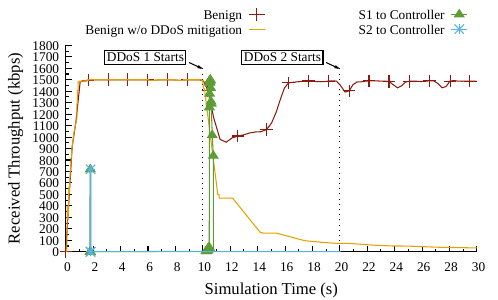}
        \label{fig:throughputResults}
    \end{subfigure}
    \hfill
    \begin{subfigure}[b]{0.44\textwidth}
        \centering
        \includegraphics[width=\linewidth]{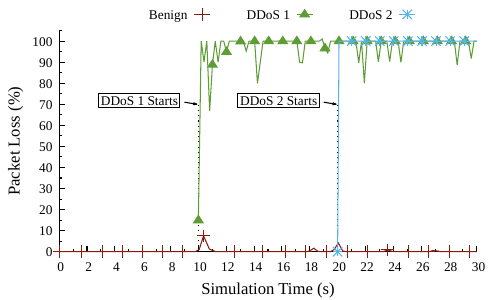}
        \label{fig:packetlossResults}
    \end{subfigure}
    \caption{Measured Network Metrics during the DDOS NS-3 simulation.}
    \label{fig:ns3_results}
\end{figure*}

Figure~\ref{fig:ns3_results} presents the results of our network-level simulation. In the baseline scenario—where no DDoS mitigation is deployed within the EP4 switches—the benign flow from C1 to C4 is rapidly overwhelmed by the malicious traffic, resulting in severe degradation of throughput. In contrast, when our proposed collaborative DDoS detection architecture is enabled, both attack phases are promptly identified and mitigated, preserving the performance of the benign flow with minimal disruption. Notably, the majority of attack flows are detected and mitigated directly within the data plane, with only a limited number of controller interactions observed. Specifically, switch-to-controller communication occurs primarily at \(t = 2\)~s and \(t = 10\)~s, corresponding to early ambiguous classifications that were escalated to the control plane or to the controller verification mechanism presented at the end of Section III. These results underscore the effectiveness of our in-network inference strategy in achieving real-time low-latency detection while significantly reducing control-plane overhead—highlighting a clear advantage over traditional controller-centric DDoS defense mechanisms.

\section{Conclusion}
This paper introduced a collaborative DDoS detection and mitigation framework that integrates P4-programmable data planes with an SDN control plane through a split early-exit neural network. By performing partial inference within the data plane and deferring low-confidence cases to the control plane, the proposed architecture enables rapid response while preserving detection accuracy.

Evaluation on real-world DDoS datasets demonstrates that the system achieves low-latency inference and reduced control plane overhead without compromising detection performance. These results underscore the effectiveness of jointly leveraging in-network processing and centralized intelligence. As future work, we aim at exploring collaborative  DDoS defense mechanisms between different P4 switches and the SDN controller, as well as supporting non-binary DDoS mitigation decisions.

\bibliographystyle{IEEEtran}
\bibliography{bibliography.bib}

\end{document}